\documentclass[preprint,3p,twocolumn,sort&compress]{elsarticle}
\usepackage[utf8]{inputenc}

\usepackage{amsmath}
\usepackage{amssymb}
\usepackage{url}
\usepackage{color}
\usepackage{siunitx}
\usepackage{booktabs}

\newcommand{\harp}{\textsc{HarP}}

\newcounter{bla}

\journal{Computer Physics Communications}

\begin{document}

\begin{frontmatter}
    \title{HarP: a simulation tool for high-order harmonic generation \\and propagation in gases}

    \author[1]{M. Pouyez \corref{author}}
       \cortext[author]{\textit{E-mail address:} mattys@chalmers.se}
    \affiliation[1]{Department of Physics and Astronomy, Chalmers University of Technology, Gothenburg, SE-41296, Sweden}
\author[1]{D. Maslarova}
\author[1]{L. Hagedorn}
\author[2]{P. Jansson}
\affiliation[2]{Department of Computer Science and Engineering, Chalmers University of Technology and University of Gothenburg, Gothenburg, Sweden}
\author[3]{R. Weissenbilder}
\affiliation[3]{Department of Physics, Lund
University, Lund, Sweden}
\author[4]{S. Carlström}
\affiliation[4]{Max Born Institute for Nonlinear Optics and Short Pulse Spectroscopy, Berlin, Germany}
\author[3]{A. L'Huillier}

\author[1]{T. Fülöp}

    \begin{abstract}
Accurately simulating high-order harmonic generation (HHG) for attosecond science requires connecting the microscopic single-atom quantum dynamics and the macroscopic propagation of intense driving laser fields. HarP is an open-source Python code designed to address this multiscale problem. It solves the wave equations for both the driving laser field and the generated harmonics, using precomputed dipole tables for the nonlinear polarization density. This tabulation enables the direct integration of accurate microscopic models while avoiding costly calculations during the propagation. 
The methodology builds upon the approach of [\textit{L'Huillier et al. (1992) PRA 46, 2778}] \cite{l1992calculations}   and has already facilitated multiple studies on phase matching, propagation effects and HHG optimization. This paper presents the underlying physical model, its numerical implementation, and comparison against alternative simulation frameworks.

    \end{abstract}

\end{frontmatter}

{\bf PROGRAM SUMMARY}\\
\begin{small}
    \noindent
    {\em Program Title:} \harp                                   \\
    {\em CPC Library link to program files:} (to be added by Technical Editor) \\
    {\em Developer's repository link:} \url{https://github.com/HarmonicPropagation/HarP} \\
    {\em Licensing provisions:} MIT\\
    {\em Programming language:}  Python                       \\
    {\em Nature of problem:} 
   Simulating and optimizing high-order harmonic generation in gases requires modelling the microscopic single-atom response and the macroscopic propagation of the generated electric field. Directly solving the three-dimensional (3D) Time-Dependent Schrödinger Equation (TDSE) on a macroscopic grid is computationally expensive, whereas computationally efficient alternatives like the Strong-Field Approximation are inherently limited in their accuracy.\\
    {\em Solution method:} HarP overcomes this computational barrier by combining macroscopic wave propagation with precomputed microscopic response tables. At the macroscopic scale, it solves the wave equations for both the driving laser and the generated harmonics under the paraxial and slowly varying envelope approximations using a Crank-Nicolson finite-difference scheme. The microscopic physics, including the atomic dipole moment and ionization rates, is obtained by interpolating externally precomputed tables. This enables the integration of accurate microscopic models such as 3D-TDSE solutions, without large computational costs during the propagation.

\end{small}

\section{Introduction}

When an intense laser pulse interacts with an atom, the electrons experience the external laser field, the Coulomb potential of the nucleus, and their own mutual interaction. The resulting electron dynamics drive various nonlinear processes, including high-order harmonic generation (HHG) \cite{mcpherson1987studies,ferray1988multiple,krause1992calculation}. 
An intuitive description of the HHG mechanism is provided by the semiclassical \textit{three-step model} \cite{corkum1993plasma,kulander1993dynamics}. 
In this model, the laser field first distorts the atomic potential, allowing a bound electron to escape through tunnel ionisation. 
The escaping electron is then accelerated by the laser field. 
Later, as the field changes direction, the electron is pulled back toward the nucleus. There is a small probability for it to recombine back to the ground state, and its excess energy is then released as an XUV photon. This process is repeated several times during the laser pulse, resulting in the emission of coherent high-order harmonic radiation.


While the three-step model provides a simple qualitative picture of the mechanism underlying HHG, an accurate description of the process requires consideration of two distinct scales \cite{gaarde2008macroscopic}.
The microscopic scale is responsible for generating the high harmonics at the atomic level and 
the macroscopic scale dictates the propagation of the harmonic \cite{heyl2017introduction}, shaped by effects such as phase matching, absorption, and dispersion, as well as the reshaping of the fundamental field by the varying electronic density.

The most accurate description of the microscopic response is obtained by solving the Time-Dependent Schrödinger Equation (TDSE) \cite{kulander1989calculations,krause1992calculation, carlstrom2022generala,carlstrom2022generalb}. 
However, due to numerical constraints, such calculations are usually limited to a single-atom response, often with the approximation of a single active electron. 
Direct treatment of harmonic propagation through the medium is therefore infeasible without additional approximations.

A widespread and computationally efficient method \cite{priori2000nonadiabatic,shon2000generation,gaarde2008macroscopic,tosa2009macroscopic,hernandez2010high,hognerHHGmax,tosa2016propagation,hogner2017generation,schroder2025open} couples macroscopic wave propagation with a microscopic response modelled via the Strong-Field Approximation (SFA).
In the SFA approach \cite{lewenstein1994theory}, the electron dynamics after ionization is assumed to be dominated by the strong laser field while the interaction with the parent ion is neglected. 
By providing a semi-analytical expression for the atomic dipole moment, the SFA-based approach allows rapid solution of the propagation equations in the frequency domain. 
Although several fast codes based on this approach are available for scanning broad parameter spaces at a minimal computational cost, their overall accuracy remains inherently limited by the SFA \cite{gaarde2002quantum,gaarde2008macroscopic}.

Another alternative has been recently developed and is referred to as the Modular Multiscale Approach (MMA) \cite{vabek2025modular,vabekMMAHHG}.
In this framework, the driving field is propagated through the target by solving the wave equation in the time domain, while the local microscopic response is computed inside each macroscopic grid point by solving the one-dimensional (1D) TDSE.
The generated high-order harmonics are subsequently propagated through the medium using a Hankel transform.
However, beyond the inaccuracies inherent to the 1D-TDSE solution compared to the three-dimensional (3D) TDSE calculation, the method's primary limitation for experimental application is its significant computational cost. 
Solving the 1D-TDSE across an entire macroscopic grid typically requires extensive high-performance computing resources.

Other open-source tools, such as Luna \cite{Brahms2021}, are suitable for more general macroscopic propagation problems that involve nonlinear processes. Luna models nonlinear optical dynamics, particularly in gas-filled waveguides and free-space geometries, by solving the unidirectional pulse propagation equation. 
It is well suited for computational optimization studies, as recently demonstrated for the Bayesian optimization of resonant dispersive-wave generation in hollow capillary fibers \cite{Klee2025}. 
While Luna is able to capture complex macroscopic propagation and optimize nonlinear optical processes, it is not designed to resolve the microscopic quantum dynamics underlying HHG. 

To overcome the computational barriers caused by the coupling of microscopic and macroscopic scales, several studies \cite{wikmark2019spatiotemporal,weissenbilder2022optimize, weissenbilder2023optimization,appi2023two, westerberg2025influence,vardast2026high} have shown that combining the framework presented in Ref.~\cite{l1992calculations} with precomputed dipole moment tables can be used to efficiently simulate and optimize HHG experiments. 
The primary advantage is that it enables the use of more accurate microscopic models, especially those relying on 3D TDSE calculations, while keeping the computational cost low. 
Nevertheless, this tabulated approach inherently relies on several simplifying approximations, which will be detailed throughout this paper.

This paper aims to introduce HarP, an open-source Python code that makes this latter approach widely accessible \cite{HarPGithub}. 
We begin in Section \ref{sec:theory} by rederiving the governing equations originally formulated in Ref. \cite{l1992calculations}. 
Section \ref{sec:numerics} describes the numerical implementation and performance metrics of the code. 
Section \ref{sec:application} presents a typical HHG simulation using HarP and compares its predictions with those of two alternative open-source frameworks: the SFA code from Ref. \cite{schroder2025open} and the MMA code from Refs. \cite{vabek2025modular,vabekMMAHHG}. 
Conclusions and future perspectives are provided in Section \ref{sec:conclusion}.

Throughout this paper, standard notation is adopted for fundamental physical constants: $c$ for the speed of light, $m_e$ for the electron mass, $e$ for the elementary charge, $\varepsilon_0$ for the vacuum permittivity, and $\hbar$ for the reduced Planck constant. The incident laser is characterized by its wavelength $\lambda$, its angular frequency $\omega=2\pi c/\lambda$, and its vacuum wavevector $k=\omega/c$.

\section{Theoretical model and approximations}
\label{sec:theory}
The theoretical foundation of the HarP code relies on the macroscopic model first described by L’Huillier et al.~\cite{l1992calculations}.
 In this section, we revisit the derivation of the governing equations, following Refs.~\cite{l1992calculations,antoine1996theory}, to explicitly highlight the model's fundamental approximations and limitations.

\subsection{Macroscopic propagation and refractive index}
We assume cylindrical symmetry with spatial coordinates  $(r,z)$, where $r$ represents the radial coordinate and $z$ is the longitudinal coordinate in the direction of propagation of the driving laser. The propagation of the $q$th harmonic electric field envelope\footnote{We have adopted the convention $\mathcal{E}(r,z,t)=1/2\sum_q \mathcal{E}_q(r,z) e^{-iq\omega t} + c.c.$ (complex conjugate) for the real electric field and $\mathcal{P}(r,z,t)=1/2\sum_q \mathcal{P}_q(r,z) e^{-iq\omega t} + c.c.$ for the nonlinear polarization.}, $\mathcal{E}_q$, is governed by the wave equation
\begin{equation}\label{eq:wave:electricfield}
    \nabla^2 \mathcal{E}_q + k_q^2 \mathcal{E}_q=-\frac{1}{\varepsilon_0} \left(\frac{q\omega}{c}\right)^2 \mathcal{P}_q.
\end{equation}
The wavevector of the $q$th harmonic is defined as $k_q=n_q q \omega/c$, with $n_q$ the refractive index of the medium evaluated at energy $q\hbar \omega$ and position $(r,z)$.
Here, $\mathcal{P}_q$ denotes the $q$th Fourier component of the nonlinear polarization.

Expressed in terms of the atomic ($\chi_a$), electronic ($\chi_e$), and ionic ($\chi_i$) susceptibilities, the refractive index reads
\begin{align}
    n_q(r,z) &= \sqrt{1+\chi_a+\chi_e+\chi_i}.
\end{align}
In what follows, we neglect the ionic susceptibility contribution with respect to the atomic and electronic ones:
\begin{align}
    \chi_a(q \omega,r,z) &= -\frac{\rho_a(r,z)}{q^2 \rho_r} \left[f_1(q\omega) -\mathbf{i} f_2(q\omega)\right], \\
    \chi_e(q \omega,r,z) &= -\frac{\rho_e(r,z)}{q^2 \rho_r}.
\end{align}
Here, $\rho_r = \varepsilon_0 m_e \omega^2 / e^2$ is the critical plasma density at the fundamental frequency, and the terms
$\rho_a$ and $\rho_e$ correspond to the atomic and electronic number densities, respectively. 
The atomic form factors\footnote{Here the atomic form factor $f_2$ is positive, consistent with the Henke tables \cite{henke1993x}.} are denoted $f_1,f_2$, which account respectively for dispersion and absorption at photon energy $q \hbar \omega$.

\subsection{Wavevector decomposition and phase mismatch}

The wavevector is expressed through a perturbative expansion: $k_q(r,z)=k_{q}^{0}(z)+\delta k_q(r,z)$.
The zeroth-order contribution $k_{q}^{0}$ is assumed to be independent of the transverse coordinate,
while the first-order perturbation $\delta k_q$ strictly satisfies the condition  $k_{q}^{0}(z) \gg \delta k_q(r,z)$.

Considering only the first ionisation relevant to the interaction, the ionic density equals the electronic density. Thus, the initial target density is $\rho_g(z)=\rho_e(r,z)+\rho_a(r,z)$, which is first assumed to be independent of the transverse coordinate. Assuming additionally a low-density target, such that $\chi_{a}, \chi_e \ll 1$, the wavevectors $\forall q>1$ take the form \cite{l1992calculations,antoine1996theory}:
\begin{align}
        \frac{c}{q\omega} k_q^0(z) &= 1 - \frac{\rho_g(z)}{2 q^2 \rho_r} \left[ f_1(q\omega) - \mathbf{i} f_2(q\omega) \right], \\
        \frac{c}{q\omega} \delta k_q(r,z) &= -\frac{\rho_e(r,z)}{2 q^2 \rho_r} \left[ 1 - f_1(q\omega) + \mathbf{i} f_2(q\omega) \right].
\end{align}
When the initial target density exhibits a radial dependence (e.g., a gas jet with a non-uniform transverse profile), one can set $k_q^0 = q\omega/c$ and absorb all spatial dependencies into the perturbation term \cite{antoine1996theory}, yielding $\delta k_q(r,z) = q\omega/c[\chi_a(q\omega,r,z) + \chi_e(q\omega,r,z)] / 2$.

For the driving optical laser ($q=1$), we define the fundamental wavevector $k_1$ in terms of the polarizability rather than the form factors. We approximate the polarizability at the driving frequency $\omega$ by its static value $\alpha_0$, which can be extracted from the literature \cite{soldan2001static,gaiser2018polarizability} or inferred from the refractive index data \cite{polyanskiy2024refractiveindex}. 

Furthermore, for lasers of high intensity (typically $\gtrsim 10^{14}$ W/cm$^2$), the Kerr effect \cite{kerr1875liv} significantly influences macroscopic propagation.
The perturbation of the fundamental field is therefore constructed to directly account for both linear polarizability and this nonlinear Kerr response \cite{boyd2008nonlinear}:
\begin{align}
    \frac{c}{\omega} k_1^0(z) &= 1 + \rho_g(z)\frac{\alpha_0}{2\varepsilon_0}, \label{eq:k1}\\
    \frac{c}{\omega} \delta k_1(r,z) &= -\frac{\rho_e(r,z)}{2} \left( \frac{\alpha_0}{\varepsilon_0} + \frac{1}{\rho_r} \right) \nonumber \\
    &\quad + \rho_a(r,z)\frac{3\gamma}{8\varepsilon_0}\lvert \mathcal{E}_1(r,z) \rvert^{2}, \label{eq:dk1}
\end{align}
where $\gamma$ denotes the second hyperpolarizability.
Note that conventions for $\gamma$ differ by a factor of~$6$ between various subfields.%
\footnote{Here, $\gamma$ is $6$ times smaller than in Ref.~\cite{soldan2001static}.}
\subsection{Slowly-varying envelope formulation of the propagation equations}

We now rewrite the wave equation of the electric field Eq.~\eqref{eq:wave:electricfield} introducing the slowly-varying envelope functions $\hat{E}_q$ and $P_q$:
\begin{align}
    \mathcal{E}_q&=\hat{E}_q \exp\left(\mathbf{i}\int^zk_{q}^{0}(z') dz'\right), \\
    \mathcal{P}_q&= P_q \exp\left(\mathbf{i}\int^z qk_{1}^{0}(z') dz'\right).
\end{align}
Considering the paraxial approximation $\partial_z^2 \hat{E}_q \ll k_{q}^{0}\partial_z \hat{E}_q$, the problem is simplified to an equation for the envelope field:
\begin{align}\label{eq:wave:eqhat}
    \nabla_\perp^2 \hat{E}_q &+ 2\mathbf{i}k_{q}^{0}(z)\partial_z \hat{E}_q +2k_{q}^{0}(z) \delta k_q(r,z) \hat{E}_q\\
    &=-\frac{1}{\varepsilon_0} \left(\frac{q\omega}{c}\right)^2 P_q \exp\left(-\mathbf{i}\int^z \Delta k_{q}^{0}(z')dz'\right), \nonumber
\end{align}
where $\Delta k_{q}^{0}(z)=k_{q}^{0}(z)-qk_{1}^{0}(z)$.
To eliminate the oscillating phase term on the right-hand side of Eq.~\eqref{eq:wave:eqhat}, we introduce a modified envelope $E_q$ defined by $\hat{E}_q = E_q \exp\left(-\mathbf{i}\int^z \Delta k_q^0(z')dz'\right)$, which yields for $q>1$:
\begin{align}
    \nabla_\perp^2 E_q + 2\mathbf{i}k_{q}^{0}(z)\partial_z E_q &+2k_{q}^{0}(z) [\delta k_q(r,z)+ \Delta k_{q}^{0}(z)]E_q \nonumber \\
    &=-\frac{1}{\varepsilon_0} \left(\frac{q\omega}{c}\right)^2 P_q, \label{eq:wave:eq}
\end{align}
and (for $q=1$):
\begin{align}
    \nabla_\perp^2 E_1 + 2\mathbf{i}k_{1}^{0}(z)\partial_z E_1 +2k_{1}^{0}(z) \delta k_1(r,z) E_1 =0.\label{eq:wave:e1}
\end{align}
Since the zeroth-order wavevector $k_1^0$ is purely real for all optical lasers, the energy calculated using the envelope $E_q$ is identical to that derived from the full electric field~$\mathcal{E}_q$.

\subsection{Microscopic nonlinear polarization}
The nonlinear polarization $P_q$, which drives high-harmonic generation, contains the microscopic response of the atom to the intense incident laser field. It is defined through the atomic dipole moment $d_q$, evaluated as the Fourier transform of the time-dependent dipole $d(t)$ oscillating at frequency $q \omega $. Under the slowly varying envelope approximation, our model assumes that $d_q$ depends strictly on the local intensity and wavelength of the driving laser. This assumption is justified because, in typical HHG experiments, the generated harmonic fields are orders of magnitude weaker than the fundamental driver. Furthermore, for a laser pulse with a sufficient number of cycles, we can assume that the strong-field atomic response depends only on the driving field intensity. Consequently, the dipole moment of the $q$th harmonic is a function of $E_1(r,z)$ only.
The nonlinear polarization therefore takes the form
\begin{equation}\label{eq:Pq}
    P_q(r,z) = 2\rho_a(r,z)d_q(E_1(r,z))e^{\mathbf{i}q\phi_1(r,z)},
\end{equation}
where $\phi_1(r,z)$ is the phase of $E_1(r,z)$. 
The factor of 2 arises from the different conventions used in the nonlinear polarisation and in the dipole moment definitions: Here, the convention $\mathcal{P}(r,z,t) = \sum_q \mathcal{P}_q(r,z) e^{-iq\omega t}/2 + c.c.$  is adopted, while the dipole $d_q$ is computed directly from the Fourier transform of $d(t)$.

The physics of HHG therefore reduces to the definition of the dipole moment. 
Because HarP evaluates this microscopic response using external, precomputed tables, a detailed derivation of the dipole moment is omitted from this section. 
A discussion regarding the microscopic models and the specific tables distributed with the current version of HarP is provided in Section \ref{sec:numerics}.

\subsection{Ionisation dynamics}
A crucial aspect of the pulse propagation is its interplay with the medium ionisation.
The probability of an atom undergoing ionisation during an infinitesimal time interval $dt$ is defined by the ionisation rate $w$.
Similar to the dipole moment, $w$ is assumed to depend solely on the local intensity and wavelength of the driving laser (ionization driven by the generated harmonics is neglected, as their intensity is typically many orders of magnitude weaker than the fundamental).
The current version of HarP considers only the first ionisation stage, and the rate $w$ is extracted from arbitrary externally computed tables.
While these tables can be generated using standard analytical frameworks like the Perelomov, Popov, Terent'ev (PPT) and Ammosov, Delone, Krainov (ADK) models \cite{perelomov1966ionization,ammosov1986tunnel}, HarP is designed with the capacity to accommodate precise microscopic descriptions.
Consequently, tabulations derived directly from 3D-TDSE solutions are preferred.
The system of equations is ultimately closed by the temporal evolution of the particle densities:
\begin{align}
     \rho_a(r,z,t) &= \rho_g(r,z)\exp\left(-\!\int^t \!w(E_1(r,z,t'))dt'\right), \label{eq:n_a}\\
        \rho_e(r,z,t) &= \rho_g(r,z)-\rho_a(r,z,t).\label{eq:n_e}
\end{align}

In these relations, the densities and electric field are explicitly time-dependent, whereas the propagation equations were previously formulated in the frequency domain. This dual treatment is justified by the two distinct time scales of the laser field: its fast optical oscillations and its slowly varying temporal envelope. Under the slowly varying envelope approximation, these time scales can be treated independently. The fast oscillations are captured in the frequency domain by decomposing the field into its harmonic components, while the $q$th field envelopes retain their explicit temporal dependence. The same separation of time scales applies to the plasma density, which should be understood as a slowly-varying function of time.

\subsection{Model assumptions}
Before detailing the numerical implementation, we outline the underlying physical assumptions of our model. These conditions define the validity limits of the current HarP framework:
\begin{itemize}
    \item[-] The interaction is restricted to a cylindrically symmetric geometry.
    \item[-] The driving laser field is assumed to be linearly polarized.
    \item[-] The spatial propagation of all fields is treated under the paraxial approximation.
    \item[-] The temporal envelopes of the fields vary slowly compared to the optical cycle.
    \item[-] The ionic susceptibility is neglected in the refractive index calculation.
    \item[-] The nonlinear polarization is induced only by the neutral atoms.
    \item[-] Nonlinear polarization and medium ionisation are driven exclusively by the fundamental laser field.
    \item[-] The atomic dipole moment depends only on the driving laser's wavelength and local intensity.
    \item[-] The initial gas target density remains far below the critical plasma density ($\varepsilon_0 m_e\omega^2/e^2$).
    \item[-] Only single ionisation is considered relevant for the interaction.
\end{itemize}

\section{Numerics and implementation}
\label{sec:numerics}

In this section, we detail the numerical procedure employed to solve the coupled system of equations \eqref{eq:wave:eq}, \eqref{eq:wave:e1}, \eqref{eq:n_a}, and \eqref{eq:n_e}. The computational space-time domain is uniformly discretized into an $N_t \times N_r \times N_z$ grid, characterized by temporal and spatial step sizes of $\Delta t$, $\Delta r$, and $\Delta z$, respectively.

\subsection{Normalisation}

All calculations are performed using dimensionless variables, employing a normalisation scheme inspired by conventions common in Particle-In-Cell (PIC) codes \cite{derouillat2018smilei}. A key advantage of this normalisation is that it simplifies the vacuum wavevector to $k_q = q$, the electronic susceptibility to $\chi_e = -\rho_e/q^2$, and the atomic susceptibility to $\chi_a = -\rho_a(f_1-\mathbf{i}f_2)/q^2$. The reference units, which are defined relative to the fundamental laser angular frequency $\omega$, are summarized in Table \ref{tab:Units}.
\begin{table}[ht]
    \centering
    \begin{tabular}{l @{\hspace{1cm}} l}
        Velocity & $V_r = c$ \\
        Time & $T_r = 1/\omega$ \\
        Length & $L_r = c/\omega$  \\
        Energy & $\mathcal{E}_r = \hbar \omega$ \\
        Density & $\rho_r=\varepsilon_0m_e\omega^2/e^2$ \\
        Electric field & $E_r=m_e c \omega/e$ \\
        Polarizability & $\alpha_r=e^2/(m_e \omega^2)$\\
        Second hyperpolarizability & $\gamma_r = e^4/(m_e^3 c^2 \omega^4)$ \\
        Dipole & $D_r=ec/\omega$
    \end{tabular}
    \caption{Normalisation units of the physical quantities.}
    \label{tab:Units}
\end{table}

\noindent Note that, although the reference time is defined by $T_r$, the temporal grid is not required to resolve this timescale. Instead, the temporal step needs to adequately sample the slowly varying envelope of the field. Similarly, the spatial grid resolution is governed by the laser waist rather than the laser wavelength.

\begin{figure}[t]
    \centering
    \includegraphics[width=0.85\linewidth]{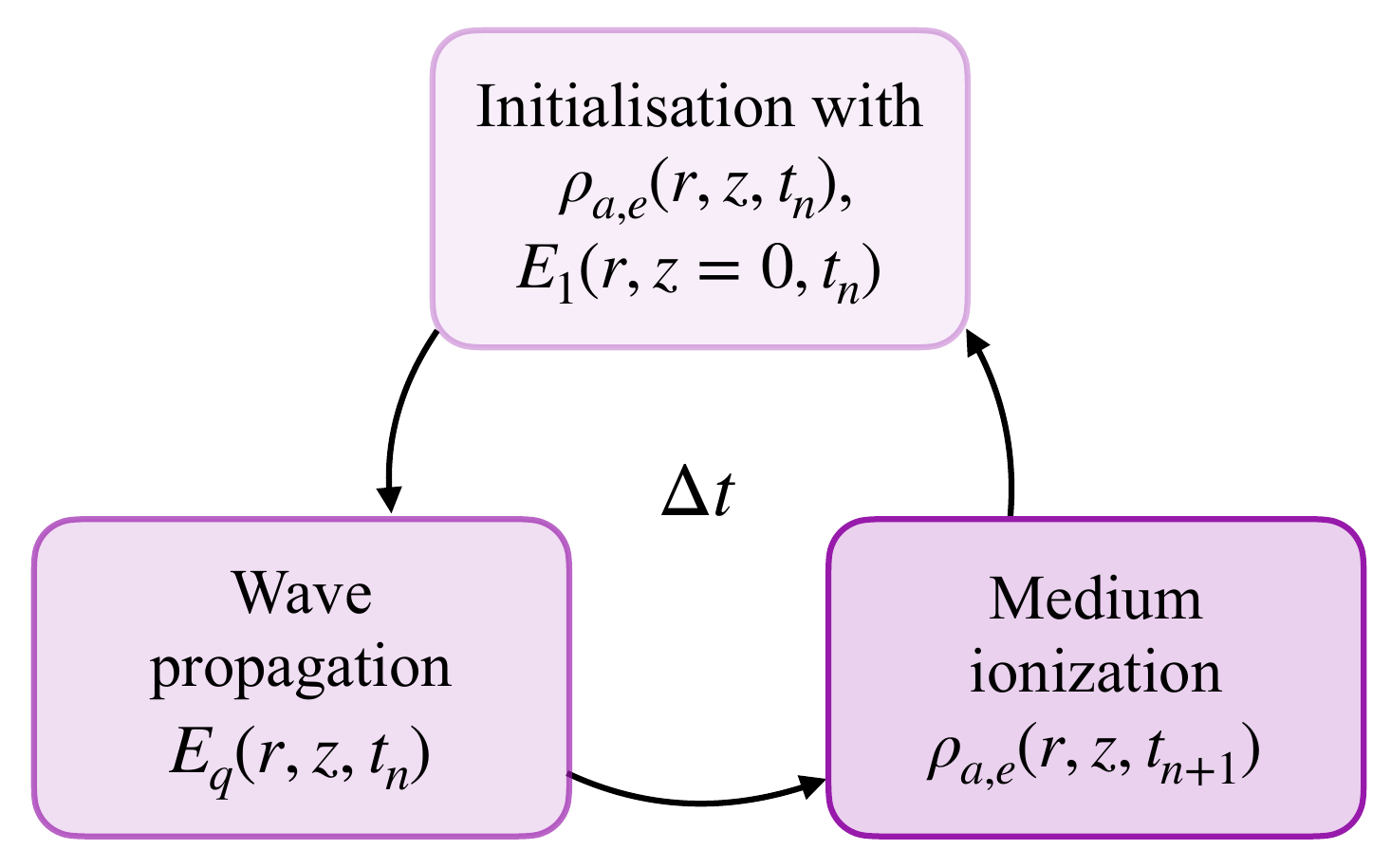}
    \caption{Schematic representation of the HarP loop. At each time slice, the propagation equations for the driver field $E_1$ and harmonic fields $E_q$ are resolved using the Crank-Nicolson scheme. Atomic and electronic densities $\rho_{a,e}$ are calculated from the driver field and used in the definition of the propagation equation at the next time step.}
    \label{fig:harp_loop}
\end{figure}
\subsection{Numerical propagation scheme}
The harmonic propagation is resolved by executing the following operations at each time step $t_n$:
\begin{enumerate}
    \item The fundamental field is initialized at the boundary with the prescribed pulse envelope, $E_1(r, z=0, t_n)$.
    \item The propagation equations \eqref{eq:wave:eq}, \eqref{eq:wave:e1} are solved over the entire $(r, z)$ spatial domain to determine the driver $E_1(r, z, t_n)$ and harmonic field envelopes $E_q(r, z, t_n)$.
    \item The atomic and electronic densities are updated according to the rate equations \eqref{eq:n_a} and \eqref{eq:n_e}. The ionisation rate is evaluated in all space from the fundamental field $E_1(r, z, t_n)$.
    \item The wave vectors and consequently, the propagation equations themselves, are updated from the new medium densities.
\end{enumerate}

\noindent A schematic representation of this numerical loop is provided in Fig.~\ref{fig:harp_loop}, and can be understood as follows: as the driving field propagates through the gas, it ionises the medium, thereby altering its own propagation. For each time slice resolving the evolution of the field envelope, the spatial propagation is computed over the entire domain. The resulting field distribution is then used to update the densities, which dictate the propagation parameters for the subsequent time iteration.

\subsection{Crank--Nicolson discretisation and boundary conditions}
The dimensionless wave equations \eqref{eq:wave:eq} and \eqref{eq:wave:e1} are solved using a Crank-Nicolson scheme \cite{crank1947practical}. Although this method is widely established, we detail the numerical procedure here for the sake of completeness. Denoting the discrete transverse and longitudinal coordinates as $(r_j,z_n)$, with $j<N_r$ and $n<N_z$, the dimensionless wave equation is recast as a second-order implicit matrix equation:
\begin{align}\label{eq:CRmatrix}
    M^{n+1}(\alpha)E^{n+1} = M^{n}(-\alpha)E^{n} + S^{n},
\end{align}
where the vector $E^{n}=(E_0^n,E_1^n,...,E_{N_r-1}^n)$ stores the electric field envelope of $r_j$ at $z_n$. Although the harmonic-order dependency ($q$) is omitted in the above notation for clarity, Eq.~\eqref{eq:CRmatrix} is solved independently for every individual harmonic and the fundamental driver field. Each possesses its own spatially discretized envelope vector $E_q$ and wavevector $k_q$. 

The tridiagonal matrix $M^n$ is composed of coefficients $m_{i,j}^n(\alpha)$ (with $\alpha=\mathbf{i}\Delta z/(4 \Delta r ^2)$) related to the wavevector denoted $k^{0,n}=k_q^{0}(z_n)$, $\Delta k^{0,n}=\Delta k^{0}_q(z_n)$ (independent of the transverse coordinate by definition) and $\delta k^{n}_j = \delta k_q(r_j,z_n)$. For internal nodes ($0 < i,j < N_r-1$), the non-zero elements are:
\begin{itemize}
    \item Lower diagonal ($j = i-1$): 
    \begin{equation*}
        m^n_{i,j}=-\dfrac{\alpha}{k^{0,n}} \left[1- \dfrac{\Delta r}{2 r_j}\right]
    \end{equation*}
    \item Main diagonal ($j = i$): 
    \begin{equation*}
        m^n_{i,j}=1 + 2\dfrac{\alpha}{k^{0,n}} - 2\alpha \left[\delta k_j^{n} + \Delta k^{0,n} \right]\Delta r^2
    \end{equation*}
    \item Upper diagonal ($j = i+1$): 
    \begin{equation*}
        m^n_{i,j}=-\dfrac{\alpha}{k^{0,n}} \left[1+ \dfrac{\Delta r}{2 r_j}\right]
    \end{equation*}
\end{itemize}

At $r=0$, a Neumann boundary condition $\partial_r E_q = 0$ is adopted. The corresponding matrix elements at $i=0$ are:
\begin{itemize}
    \item Main diagonal ($j = 0$): 
    \begin{equation*}
        m^n_{0,j}=1 + \frac{4\alpha}{k^{0,n}} - 2\alpha \left[\delta k_j^{n} + \Delta k^{0,n} \right]\Delta r^2
    \end{equation*}
    \item Upper diagonal ($j = 1$): 
    \begin{equation*}
        m^n_{0,j}=-\frac{4\alpha}{k^{0,n}}
    \end{equation*}
  \end{itemize}
At the outermost radial point, the Sommerfeld radiation condition \cite{sommerfeld1949partial} is adopted to absorb outgoing waves\footnote{To consider wave reflection, Dirichlet boundary condition $E_q(r_{max},z)=0$ is also available.}, yielding $\partial_r E_q(r_{max},z) = ik_q^{0} E_q(r_{max},z)$.
The matrix elements at $i=N_r-1$ are:
\begin{itemize}
    \item Lower diagonal ($j = N_r-2$): 
    \begin{equation*}
        m^n_{N_r-1,j}=0
    \end{equation*}
    \item Main diagonal ($j = N_r-1$): 
    \begin{equation*}
        m^n_{N_r-1,j}=1+\alpha \left[k^{0,n}-\frac{\mathbf{i}}{r_j}-2(\delta k^{n}_j + \Delta k^{0,n} )\right]\Delta r^2 
    \end{equation*}
  \end{itemize}
The vector $S^n=(S^n_0,S^n_1,...,S_{N_r-1}^n)$ is the source term responsible for the harmonic generation.
It accounts for the nonlinear polarization $P_q(r_j,z_n)=P_j^n$ term:
\begin{align}
    P_j^n = 2 \rho_{a,j}^{n} d_j^n \exp(\mathbf{i} q\phi_{1,j}^n),
\end{align}
where $\rho_{a,j}^n=\rho_a(r_j,z_n)$, $d_j^n=d_q(E_1(r_j,z_n))$ and $\phi_{1,j}^n=\arg(E_1(r_j,z_n))$, through its component definition:
\begin{align}
    S^{n}_j = q^2 \frac{\mathbf{i}\Delta z}{4}\left[\frac{P_j^{n+1}}{k^{0,n+1}}+\frac{P_j^n}{k^{0,n}}\right].
\end{align}

The electric field envelope of the driving laser is initialized at the entrance plane ($z=0$) by an arbitrary user-defined function for $E_1(r,z=0,t)$. At each time slice, the envelope at the subsequent longitudinal step $z=z_{n+1}$ is computed from the field at $z=z_n$ by solving Eq.~\eqref{eq:CRmatrix}. This is achieved using the \texttt{solve\_banded} routine from the SciPy library. This longitudinal propagation step is iterated to obtain the field $E_q$ across the entire spatial domain for each time slice.

\subsection{Kerr nonlinearity}
The Kerr effect, which is considered for the fundamental driving field only, introduces a nonlinear dependence on the electric field such that $\delta k_1(r,z) \propto \lvert E_1(r,z) \rvert^2$. To appropriately model this nonlinearity, two numerical strategies have been implemented. The most straightforward approach utilizes the Crank-Nicolson scheme, evaluating the updated wavevector $\delta k_1^{n+1}$ using the longitudinal electric field from the preceding step, $E_1^{n}$.
The second approach employs a predictor-corrector scheme. In the predictor step, an initial resolution of the matrix equation is performed by estimating the perturbation $\delta k_1^{n+1}$ from $E_1^n$. A corrector step then resolves the equation, this time evaluating both perturbations, $\delta k_1^n$ and $\delta k_1^{n+1}$, using the averaged field $(E_1^{n+1} + E_1^n)/2$. Although this iterative method doubles the computational cost for propagating the fundamental field, it improves numerical precision.

\subsection{Ionisation dynamics}

At the end of each time slice, the ionisation rate is evaluated from the fundamental field $E_1(r,z,t_n)$. The corresponding integral
\begin{align}
    Y(r,z,t_n) &= \int_0^{t_n} w(E_1(r,z,t))dt\\
    &\sim \sum_{i=0}^{n} w(E_1(r,z,t_i))\Delta t
\end{align}
is updated using a Riemann sum. 
$Y(r,z,t_n)$ is then substituted into Eqs.~\eqref{eq:n_a} and \eqref{eq:n_e} to determine the updated atomic and electronic densities.
These new densities are used to recalculate the wavevector for the next time step.

\subsection{Microscopic response tables}
The wavevectors are obtained by interpolating the atomic form factors $f_1$ and $f_2$ from external tables provided by the user. For the XUV range, widely used tabulations are available in Refs. \cite{henke1993x,HenkeTable,NistTable}. In the UV spectral range, needed for the calculation of low-order harmonics driven by IR lasers (or for less-often used UV driving lasers), the form factors may alternatively be derived from experimental measurements of the refractive index \cite{polyanskiy2024refractiveindex}.

As emphasized previously, the atomic dipole moment $d_q$ and the ionisation rate $w$ are evaluated by interpolating precomputed tables with respect to the laser intensity. Consequently, these tables must be generated specifically for the chosen driving wavelength and the atomic species under consideration.

Even though the dipole moment could be tabulated using the SFA~\cite{lewenstein1994theory} and the ionisation rate from the PPT formula~\cite{perelomov1966ionization},  we do not recommend using HarP in this manner. Although this configuration would yield results consistent with existing SFA-based propagation codes \cite{gaarde2008macroscopic,tosa2009macroscopic,hernandez2010high,hognerHHGmax,tosa2016propagation,schroder2025open}, it offers no distinct advantages in physical insight or computational performance. In fact, direct SFA-based codes are better for such cases: they bypass the slowly varying envelope approximation, which is required by our table-based approach, thereby accommodating few-cycle driving lasers.

The advantage of the tabulated approach is to incorporate complex, accurate microscopic descriptions at low computational cost. 
In contrast to recent methodologies \cite{vabek2025modular,vabekMMAHHG}, this strategy circumvents the need to perform demanding microscopic calculations at every propagation step. 
However, this computational efficiency comes at the expense of relying on precomputed data and restricting the model to the slowly varying envelope approximation.

In the current version of HarP, two precomputed tables are provided for $810$~nm lasers propagating in neon and argon gaseous media\footnote{Additional tables will be provided in future releases.}.
These tables are generated using a methodology similar to that introduced in Ref.~\cite{kulander1990generation}. 
The laser-atom interaction is simulated by solving the  3D-TDSE under the single-active-electron approximation. The driving field smoothly ramps up over five cycles and is followed by a $15$-cycle constant-amplitude plateau. 
The atomic dipole moment is Fourier transformed over the last five cycles after the transient dynamics have subsided. 
The absolute amplitude and phase are then recorded for each odd harmonic order. 
The simulation is repeated for different peak laser intensities, yielding a map of the dipole response and the ionisation rate~$w$ as a function of laser intensity and harmonic order. Because of the spin degeneracy in the single-active-electron approximation, the tabulated dipole moment and ionisation rate are a factor of 2 smaller than the ones defined in Eqs. \eqref{eq:Pq} and \eqref{eq:n_a}. This factor is taken into account when loading the tables.

Since 3D-TDSE is among the most accurate available microscopic models, we expect this tabulation to be significantly more accurate, within its regime of validity, than approaches relying on the SFA or 1D-TDSE models coupled with the PPT ionisation formula.

\subsection{Computational performance and scaling}

To efficiently handle the three-dimensional variables and minimize memory overhead, the harmonic field envelopes and densities are stored as vectors of size $N_r$, which contain the radial distribution at a given longitudinal position and time step.
These vectors are dynamically updated throughout the HarP execution loop and saved at user-specified intervals via external diagnostic classes.
A description of the code outputs is available in the HarP documentation \cite{HarPGithub}, and examples are presented in the following sections.

The computational complexity of the algorithm scales linearly with the total number of space-time grid points and the number of harmonics $N_q$, yielding a time complexity of $\mathcal{O}(N_q N_r N_z N_t)$ (see \ref{app:computationalscaling}).
As a practical example, running the code on a standard laptop (Apple MacBook Air M5, 16 GB RAM) requires a total simulation time of approximately $6 \times (N_q +1) \times (N_r/512) \times (N_z/256) \times (N_t/256)$ seconds. 

Because the spatial grid is marched longitudinally ($z$) and temporally ($t$), the full four-dimensional grid does not need to be held in memory simultaneously.

The working set memory footprint at any given step is bounded strictly by the harmonic fields in the radial dimension alongside the atomic and electronic densities tracked across the longitudinal and radial dimensions.
This yields an efficient space complexity of $\mathcal{O}(N_q N_r + N_z N_r)$.
This localized memory requirement prevents out-of-memory errors on consumer hardware, enabling high-resolution parameter scans without relying on high-performance computing clusters.

\subsection{Software design and validity checks}
From a software engineering perspective, HarP has a modular, object-oriented Python architecture:
the macroscopic wave propagation solver is a separate class from the microscopic response interpolators (dipole moment and ionisation rate).
Thus, substituting or upgrading a table for a different atomic target or theoretical model does not require touching the propagation code.

Reliability and reproducibility are checked by a GitHub continuous integration (CI) pipeline (for Ubuntu and macOS, for Python 3.13 and 3.14). 
At each commit the pipeline runs static type checking (mypy) and a pytest test suite covering component correctness, the runtime assumption checks described below, physics sanity checks against known analytical results, and optionally performance/scaling benchmarks.

Two of the modelling assumptions from Sec.~\ref{sec:theory} can be verified automatically at the start of each run: the slowly-varying-envelope condition and the low-density condition, both raising a warning if violated.
Some of the remaining assumptions (e.g. the cylindrical symmetry and linear polarisation of the driving field) are structural properties of the current implementation rather than conditions that could be violated at runtime. 
Others, notably the paraxial approximation and the degree of ionisation, are genuine dynamical conditions that can only be assessed as the simulation evolves. 
While a dynamic check for the paraxial approximation is implemented, user-enabled runtime checks for the other limitations could be considered in future releases.

\section{Application and comparison with alternative frameworks}\label{sec:application}

\begin{figure*}[t]
    \centering
    \includegraphics[width=1\linewidth]{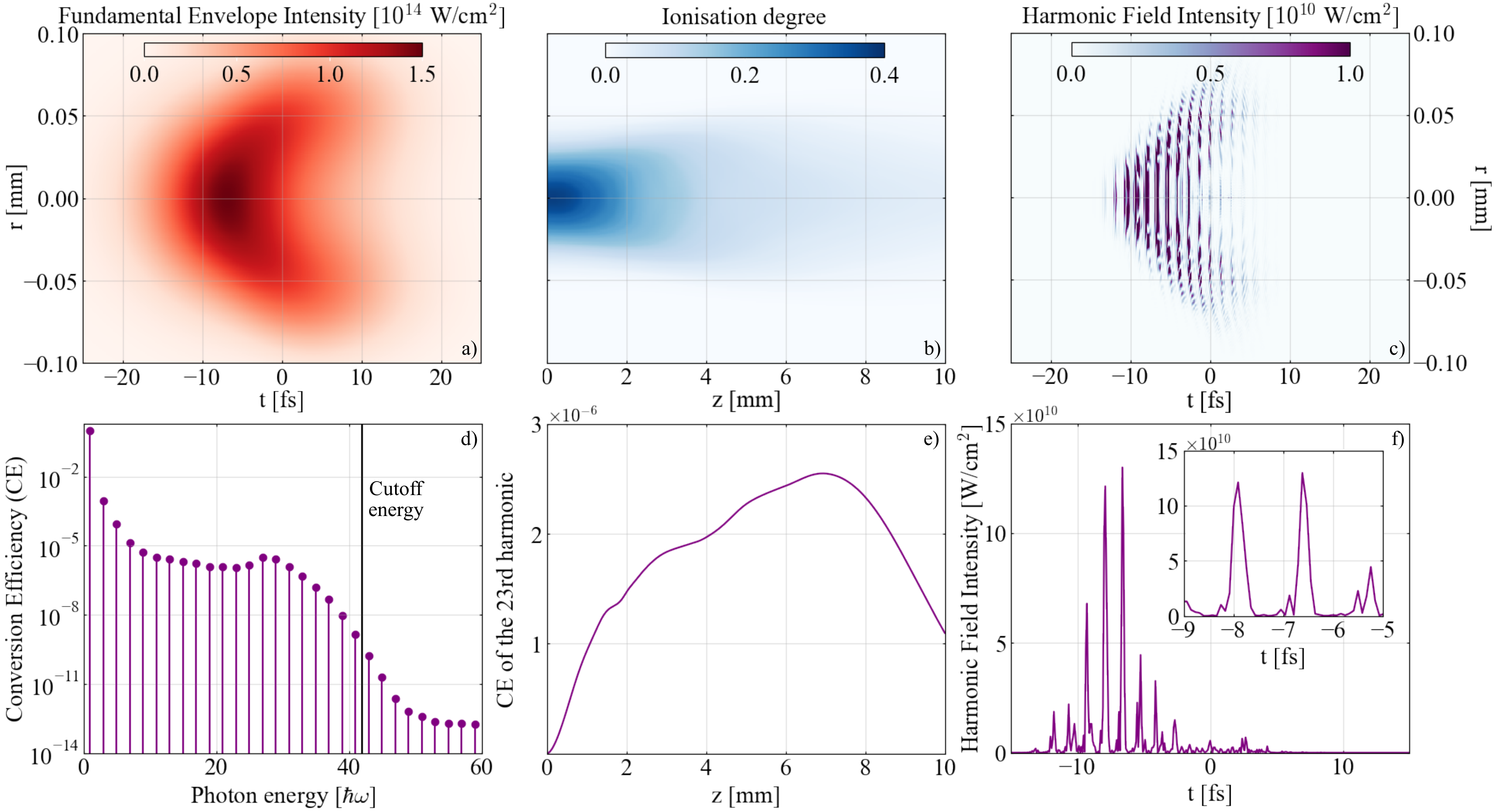}
    \caption{Examples of quantities extracted from the HarP simulation. 
    The simulation parameters are described in the main text. 
    Panel (a): the fundamental envelope intensity just after the target as a function of time and $r$. 
    Panel (b): the ionisation degree at the end of the interaction (electronic density over the initial gas density). 
    Panel (c): the reconstructed harmonic field intensity at the end of the target. 
    It has been obtained considering all the harmonics of energy larger than $35$ eV, equivalently all odd harmonics in the range $q \in [23, 75]$. 
    The same quantity at $r=0$ is presented as a function of time in panel (f). The inset is a zoom in the region $[-9,-5]$ fs. 
    Panel (d): the conversion efficiency at the end of the target as a function of the photon energy. The black solid line stands for the cutoff energy Eq. \eqref{eq:Ecutoff}.
    Panel (e): the conversion efficiency of the $23$rd harmonic as a function of the propagation distance.}
    \label{fig:allplot}
\end{figure*}

The simulations were performed using HarP version 1.0.0 available in the active public GitHub repository \cite{HarPGithub}. The input files used in this section are archived in Zenodo \cite{HarPZenodo}.

To illustrate the capabilities of HarP, we present a representative HHG simulation using parameters corresponding to an optimal low-pressure-length configuration identified in \cite{weissenbilder2022optimize}.
The driving laser pulse is assumed to have Gaussian temporal and spatial profiles,
with a wavelength of $810$ nm,
a beam waist of $w_0=71.8$ $\mu$m,
corresponding to a Rayleigh length of $z_R=20$ mm,
and a temporal intensity full width at half maximum of $T_{\rm FWHM}=22$ fs.
The peak intensity in vacuum is $I_0=2.5\times10^{14}$ W/cm$^2$. 
The laser is focused at the centre of an argon gas jet with a uniform transverse and longitudinal density profile. 
The target spans a length of $L=10$ mm and an atomic density of $\rho_0=4.9\times10^{17}$~cm$^{-3}$, corresponding to a pressure of $20$ mbar at $293$~K.
The density profile is initially uniform throughout space. 
The atomic form factors at XUV energies are defined from the Henke tables \cite{HenkeTable,henke1993x}, which provide $f_1$ for energies above $29$ eV and $f_2$ for energies above $15$ eV. At energies below $29$ eV, $f_1$ is instead computed from the phase mismatch reported in Ref.~\cite{l1990propagation} while at energies below $15$ eV, absorption is neglected, setting $f_2$ to zero.

The simulation domain extends from $r=0$ to $4w_0$ in the radial direction, from $z=0$ to $L$ along the propagation direction, and encompasses a temporal window of $4T_{\rm FWHM}$.
The domain is discretized using $N_r=512$,
$N_z=256$, and $N_t=1024$ grid points\footnote{In general, such time resolution is not necessary to accurately compute the energy and envelope field evolution. Discretising the temporal window of $4 T_{\rm FWHM}$ over $N_t=256$ points is generally sufficient for these observables. However, we choose $N_t=1024$ here to illustrate the attosecond structure of the reconstructed field (Fig.~\ref{fig:allplot}(c),(f)).}. Harmonic propagation is calculated for odd harmonic orders in the range $q\in[1,75]$. The simulation took $\sim 15$ min on an Apple MacBook Air M5, 16 GB RAM.

Figure~\ref{fig:allplot}(a) shows the cycle-averaged intensity of the fundamental field $\varepsilon_0c \lvert \mathcal{E}_1 \rvert ^2/2$ at the exit of the gas target as a function of time and radial position. 
As the driving field propagates through the medium, it ionises the gas, thereby modifying the electronic and atomic densities. 
The resulting plasma contribution changes the refractive index and consequently affects the propagation of the fundamental field. 
This nonlinear propagation leads to a deviation from the initial Gaussian profile, as observed in Fig.~\ref{fig:allplot}(a).  
The corresponding ionisation degree distribution, equivalently the electron density normalized to the initial gas density $\rho_0$, at the end of the interaction is shown in Fig.~\ref{fig:allplot}(b). Since the field has not yet been dispersed at the target entrance, it remains more intense and induces a higher ionisation in this region.

The real harmonic electric field in time after an arbitrary filter placed at the end of the target reads:
\begin{equation}\label{eq:TotalElectricField}
    \mathcal{E}(r,z,t)=\frac{1}{2}\sum_{q\in[q_1,q_2]} \mathcal{E}_q(r,z,t)e^{-\mathbf{i}q\omega t} + c.c.,
\end{equation}
where $q_1$ and $q_2$ represent the harmonic order interval of the filter. To highlight the attosecond structure of the generated emission, Fig.~\ref{fig:allplot}(c) displays the instantaneous intensity of this reconstructed field $I(r,t)=\varepsilon_0 c \lvert \mathcal{E}(r,z=L,t) \rvert^2$ at the target exit ($z=L$) assuming a high-pass filter starting at $q_1=23$ ($q_1 \hbar \omega \approx 35$~eV) and ending at $q_2=75$ (the last computed harmonic).
The corresponding temporal profile along the propagation axis ($r=0$) is presented in Fig.~\ref{fig:allplot}(f).

\begin{figure*}[t]
    \centering
    \includegraphics[width=1\linewidth]{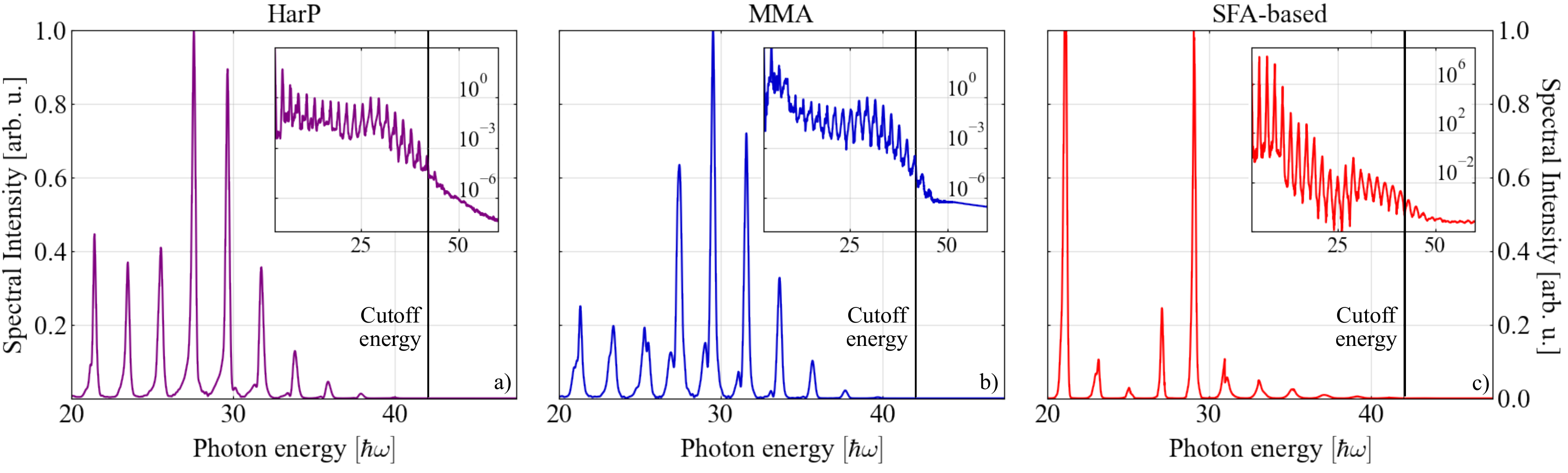}
    \caption{Comparison of the HHG spectral intensity obtained with HarP, SFA-based propagation, and MMA as a function of photon energy, equivalently, harmonic order. 
    The simulation parameters are described in the main text. 
    Panel (a) shows the result obtained with HarP \cite{HarPGithub}, 
    panel (b) the result from the MMA code \cite{vabek2025modular,vabekMMAHHG}, and 
    panel (c) the result from the SFA-based propagation code \cite{schroder2025open}.
    In each panel, the black solid line indicates the theoretical cutoff energy given by Eq.~\eqref{eq:Ecutoff}. 
    The insets show the corresponding spectra on a logarithmic $y$-axis scale. 
    Each spectrum is normalised by an arbitrary factor so the respective plateaus have comparable magnitude.
}
    \label{fig:compare_codes}
\end{figure*}

From the field envelopes, the conversion efficiency (CE) can be evaluated as a function of the propagation distance $z$. Introducing the total harmonic energy $U_q$ as a function of the propagation distance:
\begin{equation}\label{eq:Energy}
    U_q(z)= \pi \varepsilon_0 c \int_{-\infty}^{\infty} dt \int_{0}^{\infty} dr \lvert E_q (r,z,t) \rvert^2 r \,.
\end{equation}
The CE is defined as:
\begin{align}\label{eq:CE}
    \text{CE}_{q}(z) = \frac{U_q(z)}{U_1(z)}, 
\end{align}
where the denominator corresponds to the energy of the incident field, which is conserved along the propagation since there are no absorption terms in the wavevector $k_1$ Eqs. \eqref{eq:k1} and \eqref{eq:dk1}. The harmonic energy is computed through a dedicated HarP diagnostic. The CE at the exit of the gas medium is shown\footnote{Here we highlight the CE only, but the total harmonic energy provides the same behaviour.} in Fig.~\ref{fig:allplot}(d) as a function of photon energy. 
Since the photon energy is expressed in units of $\hbar\omega$, the horizontal axis can equivalently be interpreted in terms of harmonic order. 
The cutoff energy induced by a laser field of strength $E_0$ and predicted by the three-step model \cite{corkum1993plasma,kulander1993dynamics} reads:
\begin{equation}\label{eq:Ecutoff} U_{\rm{cutoff}} = I_p +3.17 \frac{e^2 \lvert E_0 \rvert^2 }{4 m_e \omega^2} \,. 
\end{equation}
For the parameters considered here, $I_p \sim 15.75$ eV, $E_0\sim 4.3 \times 10^{10}$ V/m, yielding $U_{\rm{cutoff}}\sim 42\hbar\omega \sim 64$ eV. 
It is indicated by the black solid line. 
As shown in the figure, the simulation reproduces the three characteristic regimes of HHG, namely, the perturbative regime, the plateau, and the cutoff. 
The theoretical cutoff energy Eq.~\eqref{eq:Ecutoff} is in excellent agreement with the end of the plateau.

The evolution of the CE of the $23$rd harmonic is presented as a function of $z$ in Fig.~\ref{fig:allplot}(e).
The variation of the CE with the propagation distance arises from the competition among harmonic generation, propagation, and absorption. 
As finding the optimal target configuration \cite{rundquist1998phase,schnurer1999absorption,constant1999optimizing,kazamias2003global,he2009spatial,rothhardt2014absorption,weissenbilder2022optimize,weissenbilder2023optimization,attosecond_roadmap} is an active subject of research that mainly requires maximising the CE, HarP is designed to enable its efficient evaluation and optimization.

Another important experimental observable is the HHG spectrum $S(\omega)$. 
While this quantity is not directly computed in the HarP simulation, it can be reconstructed in post-processing. 
First, the total electric field is computed using Eq.~\eqref{eq:TotalElectricField} with $q_1=1$ and $q_2=75$ the last computed harmonic. 
The resulting field is Fourier transformed and integrated over the transverse coordinate, yielding the spectral intensity (in arbitrary units) at the target exit:
\begin{equation}\label{eq:PowerSpectrum}
    S(\omega)= \int_0^{\infty} \left\lvert \mathcal{E}(r,z=L,\omega)\right\rvert^2  r dr .
\end{equation}
However, accurate resolution of the high-frequency
components in Fourier space requires the simulation to resolve the attosecond timescale. Therefore,
a second simulation was performed using the same laser–gas configuration, but with an increased temporal resolution of $N_t=16384$ and an extended temporal window of $5T_{\mathrm{FWHM}}$. This second simulation took $\sim 4$ hours instead. The resulting harmonic energy\footnote{The harmonic energy (equivalently CE Fig.~\ref{fig:allplot}(d)) could also be computed from the spectral intensity by integrating it in the neighborhood of each odd harmonic. As the simulation directly solves the problem in time, integration with Eq. \eqref{eq:Energy} is naturally preferred.} (equivalently CE) and electric profile are the same as those presented in Fig. \ref{fig:allplot}, but this last simulation allows, in addition, to accurately compute the spectral intensity according to Eq. \eqref{eq:PowerSpectrum}. 

The resulting HHG spectrum, normalised to its maximum amplitude for photon energies above $20\hbar\omega$, is shown in Fig.~\ref{fig:compare_codes}(a). 
The normalisation is chosen such that the plateau region is of order unity, allowing the spectra of the different approaches to be compared on a similar scale.
The inset shows the same spectrum on a logarithmic $y$-axis scale. The black solid line indicates the theoretical cutoff energy given by the three-step model Eq.~\eqref{eq:Ecutoff}.

For comparison, Figures~\ref{fig:compare_codes}(b) and (c) present the HHG spectra generated by the MMA code detailed in Refs.~\cite{vabek2025modular,vabekMMAHHG}, and the SFA-based propagation code implemented in Ref.~\cite{schroder2025open}, respectively. Both simulations employ the identical physical configuration described previously.
The input parameters and files for these simulations are archived on Zenodo \cite{HarPZenodo} rather than being explicitly detailed here. The SFA-based propagation simulation was performed in tens of minutes on a standard laptop, whereas the MMA simulation required several hundred CPU hours on a supercomputer.

While the HHG spectrum is directly provided by the SFA-based output \cite{schroder2025open}, for the MMA simulation it was obtained by Fourier transforming the final electric field and integrating it over the transverse coordinate, similarly to Eq.~\eqref{eq:PowerSpectrum}.

In the MMA simulations, the atomic form factors are obtained either from the Henke tables or the NIST tables, which both are not defined in the low-photon-energy regime $\lesssim 20\hbar\omega$. 
To obtain the spectrum over the full energy range, the MMA code was modified to use the same form factors as those employed in the HarP simulations, incorporating the low-energy form factor computed from the phase mismatch reported in Ref.~\cite{l1990propagation}.
This adjustment enables the calculation of the spectrum for photon energies below $\sim20\hbar\omega$.

In the SFA-based simulation, the dispersion and absorption are accounted for in the complex refractive index $n=1-\delta-\mathbf{i}\beta$. 
The parameters $\delta(q)=\rho_0/(2q^2\rho_r )f_1$ and $\beta(q)=-\rho_0/(2q^2\rho_r )f_2$ are tabulated from the same atomic form factors used in the HarP simulation.

As illustrated in Fig.~\ref{fig:compare_codes}, the results obtained with the HarP and MMA simulations show excellent agreement in their trends, in particular for the peak positions, the cutoff regime, as well as the plateau shape and its energy range. However, in the SFA-based simulation, the perturbative and the plateau regimes extend over different energy ranges with respect to the two other approaches. 
The quantitative differences in these three simulations can primarily be attributed to the microscopic models used for the atomic dipole: the tabulated approach combined with the 3D-TDSE model in HarP, the SFA model, and the 1D-TDSE microscopic response used in MMA. 
Each approach balances physical accuracy against computational efficiency. The purpose of this comparison is not to single out one superior model. 
Instead, it shows how different microscopic descriptions shift the predicted HHG spectrum and, with it, the optimal HHG configuration.
The physical origin and implications of
these discrepancies are left for future work.

\section{Conclusions}
\label{sec:conclusion}
In summary, HarP is a fast compact computational framework for solving high-order harmonic generation in gases. It couples macroscopic propagation equations with precomputed microscopic response tables, thereby incorporating accurate microscopic physics without the computational burden of solving the TDSE during propagation.
Beyond standard uniform gas jets, HarP can accommodate tailored complex target shapes, such as capillary waveguides \cite{durfee1999phase}. 
The modular design also facilitates the integration of new atomic species by substituting the externally precomputed dipole and ionization tables.  
The framework allows users to change the driving field parameters, including the application of custom temporal or spatial envelopes and varying spatial focus positions. 
HarP is therefore a flexible tool for designing, analysing, and optimizing attosecond light sources. 

The methodology presented in this work has already been successfully applied to several high-harmonic generation studies \cite{wikmark2019spatiotemporal,weissenbilder2022optimize, weissenbilder2023optimization,appi2023two, westerberg2025influence,vardast2026high}, demonstrating its potential for the simulation and optimization of HHG experiments. 
In particular, it has been used to determine optimal gas-jet lengths and pressures for a given driving laser \cite{weissenbilder2022optimize}, as well as to identify optimal laser intensities and pulse durations for specific gas targets \cite{vardast2026high}. 
Building on these multidimensional parameter dependencies, HarP has recently been integrated into a physics-informed Bayesian optimization framework \cite{Dominikaspaper}. 
By using the solver to map out phase-matching conditions across several dimensions, this approach substantially speeds up the search for high-efficiency, robust operating regimes.
This work makes that methodology available to the community through HarP \cite{HarPGithub}, an open-source Python package.

Additional microscopic response tables covering different atomic species and laser wavelengths will become available in the future. Further developments are under investigation to increase the flexibility of the microscopic response and extend the range of propagation conditions. This includes taking into account ionic susceptibility, double ionisation, elliptically polarised lasers, non-uniform mesh grids, other potential diagnostics, solver and boundary conditions.

HarP is designed with experimental optimization in mind, where maximising the conversion efficiency is essential. The combination of an accurate microscopic response with a fast macroscopic propagation model makes such parameter studies accessible on standard computing hardware, without requiring dedicated high-performance computing resources for each simulation. Complete documentation, examples, and information on the use and development of the code are provided with the open-source distribution \cite{HarPGithub}.

\section*{Acknowledgements}
The authors gratefully acknowledge fruitful discussions with Filip Salbrink, with Frederic Perez and Mickael Grech on the code documentation, and with Jan Vabek on the MMA simulations.

\section*{Funding}
This project received funding from the Knut and Alice Wallenberg Foundation (Grant Nos.~KAW 2020.0111, 2023.0249, 2023.0207) and from the Swedish Research Council (Grant No.~2023-04603). A.~L.~acknowledges support from the Knut and Alice Wallenberg Foundation through the Wallenberg Center for Quantum Technology. The computations were enabled by resources provided by the National Academic Infrastructure for Supercomputing in Sweden (NAISS), partially funded by the Swedish Research Council through grant agreement No. 2022-06725.

\section*{Competing Interests}
The authors declare no competing interests.

\appendix

\section{Computational scaling}
\label{app:computationalscaling}
Figure~A.4 illustrates the computational scaling of HarP as a function of the numerical grid resolution and the number of propagated harmonic orders. The results confirm the expected linear dependence of the simulation time on each of the discretization parameters, $N_r$, $N_z$, and $N_t$, as well as on the number of harmonics $N_q$. When varying one parameter, all remaining grid dimensions are fixed to $N_r=N_z=N_t=256$ and $N_q=1$. The measured computation times closely follow the complexity estimate given by
\begin{equation}
    6(N_q+1)
    \left(\frac{N_r}{512}\right)
    \left(\frac{N_z}{256}\right)
    \left(\frac{N_t}{256}\right)
    \ \mathrm{s},
\end{equation}
confirming the approximately linear scaling predicted by the computational complexity $\mathcal{O}(N_q N_r N_z N_t)$. This favourable scaling makes it possible to perform high-resolution simulations and parameter scans on standard computing hardware.
\begin{figure}[ht]
    \centering
    \includegraphics[width=0.9\linewidth]{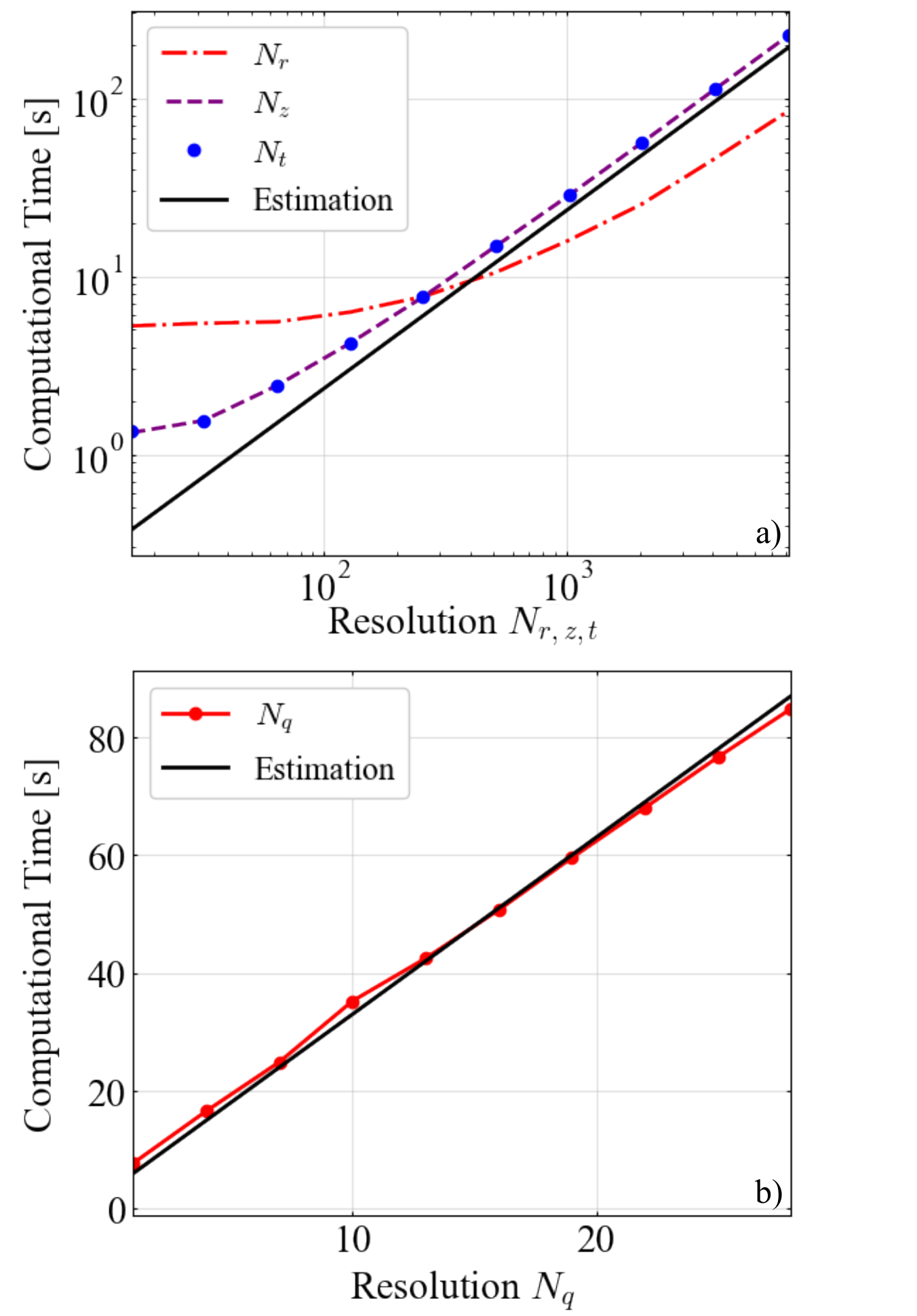}
    \caption{Computational time as a function of the grid resolution. Panel (a): as a function of $N_r$ in red dashdotted line, $N_z$ in purple dashed line, and $N_t$ in blue dots. When varying $N_r$, $N_t$ or $N_z$, all the other are set to $256$, and $N_q=1$. Panel (b): as a function of $N_q$ in red, with $N_r=N_z=N_t=256$. In both panels, the black solid line represents the complexity estimate $6 \times (N_q +1) \times (N_r/512) \times (N_z/256) \times (N_t/256)$ seconds. }
    \label{fig:placeholder}
\end{figure}


\begin{thebibliography}{10}
\expandafter\ifx\csname url\endcsname\relax
  \def\url#1{\texttt{#1}}\fi
\expandafter\ifx\csname urlprefix\endcsname\relax\def\urlprefix{URL }\fi
\expandafter\ifx\csname href\endcsname\relax
  \def\href#1#2{#2} \def\path#1{#1}\fi

\bibitem{l1992calculations}
A.~L’Huillier, P.~Balcou, S.~Candel, K.~J. Schafer, K.~C. Kulander, Calculations of high-order harmonic-generation processes in xenon at 1064 nm, Physical Review A 46~(5) (1992) 2778.

\bibitem{mcpherson1987studies}
A.~McPherson, G.~Gibson, H.~Jara, U.~Johann, T.~S. Luk, I.~McIntyre, K.~Boyer, C.~K. Rhodes, Studies of multiphoton production of vacuum-ultraviolet radiation in the rare gases, Journal of the Optical Society of America B 4~(4) (1987) 595--601.

\bibitem{ferray1988multiple}
M.~Ferray, A.~L'Huillier, X.~Li, L.~Lompre, G.~Mainfray, C.~Manus, Multiple-harmonic conversion of 1064 nm radiation in rare gases, Journal of Physics B: Atomic, Molecular and Optical Physics 21~(3) (1988) L31--L35.

\bibitem{krause1992calculation}
J.~L. Krause, K.~J. Schafer, K.~C. Kulander, Calculation of photoemission from atoms subject to intense laser fields, Physical Review A 45~(7) (1992) 4998.

\bibitem{corkum1993plasma}
P.~B. Corkum, Plasma perspective on strong field multiphoton ionization, Physical review letters 71~(13) (1993) 1994.

\bibitem{kulander1993dynamics}
K.~Kulander, K.~Schafer, J.~Krause, Dynamics of short-pulse excitation, ionization and harmonic conversion, in: Super-intense laser-atom physics, Springer, 1993, pp. 95--110.

\bibitem{gaarde2008macroscopic}
M.~B. Gaarde, J.~L. Tate, K.~J. Schafer, Macroscopic aspects of attosecond pulse generation, Journal of Physics B: Atomic, Molecular and Optical Physics 41~(13) (2008) 132001.

\bibitem{heyl2017introduction}
C.~Heyl, C.~L. Arnold, A.~Couairon, A.~L’Huillier, Introduction to macroscopic power scaling principles for high-order harmonic generation, Journal of Physics B: Atomic, Molecular and Optical Physics 50~(1) (2017) 013001.

\bibitem{kulander1989calculations}
K.~Kulander, B.~Shore, Calculations of multiple-harmonic conversion of 1064-nm radiation in xe, Physical review letters 62~(5) (1989) 524.

\bibitem{carlstrom2022generala}
S.~Carlstr{\"o}m, M.~Spanner, S.~Patchkovskii, General time-dependent configuration-interaction singles. i. molecular case, Physical Review A 106~(4) (2022) 043104.

\bibitem{carlstrom2022generalb}
S.~Carlstr{\"o}m, M.~Bertolino, J.~M. Dahlstr{\"o}m, S.~Patchkovskii, General time-dependent configuration-interaction singles. ii. atomic case, Physical Review A 106~(4) (2022) 042806.

\bibitem{priori2000nonadiabatic}
E.~Priori, G.~Cerullo, M.~Nisoli, S.~Stagira, S.~De~Silvestri, P.~Villoresi, L.~Poletto, P.~Ceccherini, C.~Altucci, R.~Bruzzese, et~al., Nonadiabatic three-dimensional model of high-order harmonic generation in the few-optical-cycle regime, Physical Review A 61~(6) (2000) 063801.

\bibitem{shon2000generation}
N.~H. Shon, A.~Suda, K.~Midorikawa, Generation and propagation of high-order harmonics in high-pressure gases, Physical Review A 62~(2) (2000) 023801.

\bibitem{tosa2009macroscopic}
V.~Tosa, K.~T. Kim, C.~H. Nam, Macroscopic generation of attosecond-pulse trains in strongly ionized media, Physical Review A—Atomic, Molecular, and Optical Physics 79~(4) (2009) 043828.

\bibitem{hernandez2010high}
C.~Hern{\'a}ndez-Garc{\'\i}a, J.~A. P{\'e}rez-Hern{\'a}ndez, J.~Ramos, E.~C. Jarque, L.~Roso, L.~Plaja, High-order harmonic propagation in gases within the discrete dipole approximation, Physical Review A—Atomic, Molecular, and Optical Physics 82~(3) (2010) 033432.

\bibitem{hognerHHGmax}
M.~Högner, V.~Yakovlev, I.~Pupeza, {HHGmax}: Simulation framework for high-harmonic generation, \url{https://gitlab.com/mhoegner/HHGmax} (2015).

\bibitem{tosa2016propagation}
V.~Tosa, K.~Kov{\'a}cs, B.~Major, E.~Balogh, K.~Varj{\'u}, Propagation effects in highly ionised gas media, Quantum Electronics 46~(4) (2016) 321--326.

\bibitem{hogner2017generation}
M.~H{\"o}gner, V.~Tosa, I.~Pupeza, Generation of isolated attosecond pulses with enhancement cavities—a theoretical study, New Journal of Physics 19~(3) (2017) 033040.

\bibitem{schroder2025open}
C.~A. Schr{\"o}der, R.~Kienberger, S.~R. Leone, Open-source simulation program for extreme ultraviolet and soft x-ray sources based on high-harmonic generation, Optics Express 33~(23) (2025) 48214--48226.

\bibitem{lewenstein1994theory}
M.~Lewenstein, P.~Balcou, M.~Y. Ivanov, A.~L’huillier, P.~B. Corkum, Theory of high-harmonic generation by low-frequency laser fields, Physical Review A 49~(3) (1994) 2117.

\bibitem{gaarde2002quantum}
M.~B. Gaarde, K.~J. Schafer, Quantum path distributions for high-order harmonics in rare gas atoms, Physical Review A 65~(3) (2002) 031406.

\bibitem{vabek2025modular}
J.~V{\'a}bek, T.~N{\v{e}}mec, S.~Skupin, F.~Catoire, Modular multiscale approach to modelling high-harmonic generation in gases, arXiv preprint arXiv:2507.04115 (2025).

\bibitem{vabekMMAHHG}
J.~Vábek, T.~Němec, S.~Skupin, F.~Catoire, {Multiscale modular approach (MMA) for HHG modelling}, \url{https://mma-hhg.github.io} (2026).

\bibitem{Brahms2021}
C.~Brahms, J.~C. Travers, \href{https://github.com/LupoLab/Luna.jl}{Luna.jl: A flexible nonlinear optical pulse propagator} (2021).
\newblock \href {https://doi.org/10.5281/zenodo.5513570} {\path{doi:10.5281/zenodo.5513570}}.
\newline\urlprefix\url{https://github.com/LupoLab/Luna.jl}

\bibitem{Klee2025}
T.~Klee, R.~Mukherjee, J.~J. Broughton, C.~Ferchaud, C.~Brahms, J.~C. Travers, F.~Mintert, J.~W.~G. Tisch, Bayesian optimization of resonant dispersive wave generation in hollow capillary fibers, Optics Express 33 (2025) 7027.
\newblock \href {https://doi.org/10.1364/oe.547228} {\path{doi:10.1364/oe.547228}}.

\bibitem{wikmark2019spatiotemporal}
H.~Wikmark, C.~Guo, J.~Vogelsang, P.~W. Smorenburg, H.~Coudert-Alteirac, J.~Lahl, J.~Peschel, P.~Rudawski, H.~Dacasa, S.~Carlstr{\"o}m, et~al., Spatiotemporal coupling of attosecond pulses, Proceedings of the National Academy of Sciences 116~(11) (2019) 4779--4787.

\bibitem{weissenbilder2022optimize}
R.~Weissenbilder, S.~Carlstr{\"o}m, L.~Rego, C.~Guo, C.~Heyl, P.~Smorenburg, E.~Constant, C.~Arnold, A.~L’huillier, How to optimize high-order harmonic generation in gases, Nature Reviews Physics 4~(11) (2022) 713--722.

\bibitem{weissenbilder2023optimization}
R.~Weissenbilder, Optimization of high-order harmonic generation for attosecond science, no. 593, Lund University, 2023.

\bibitem{appi2023two}
E.~Appi, R.~Weissenbilder, B.~Nagyill{\'e}s, Z.~Diveki, J.~Peschel, B.~Farkas, M.~Plach, F.~Vismarra, V.~Poulain, N.~Weber, et~al., Two phase-matching regimes in high-order harmonic generation, Optics Express 31~(20) (2023) 31687--31697.

\bibitem{westerberg2025influence}
S.~Westerberg, M.~Redon, A.-K. Raab, G.~Beaufort, M.~Arias~Velasco, C.~Guo, I.~Sytcevich, R.~Weissenbilder, D.~O’Dwyer, P.~Smorenburg, et~al., Influence of the laser pulse duration in high-order harmonic generation, APL Photonics 10~(9) (2025).

\bibitem{vardast2026high}
S.~Vardast, A.~Muschet, N.~Smijesh, M.~Rezaei-Pandari, F.~Schnur, R.~Weissenbilder, E.~Appi, J.~Lahl, S.~Maclot, P.~Eng-Johnsson, et~al., High-intensity attosecond beamline for xuv pump--xuv probe measurements with photon energies up to 150 ev, APL Photonics 11~(5) (2026).

\bibitem{HarPGithub}
M.~Pouyez, D.~Maslarova, L.~Hagedorn, P.~Jansson, R.~Weissenbilder, S.~Carlstr{\"o}m, A.~L’Huillier, T.~F{\"u}l{\"o}p, {HarP: a simulation tool for high harmonic propagation in gases.}, \url{https://github.com/HarmonicPropagation/HarP} (2026).

\bibitem{antoine1996theory}
P.~Antoine, A.~L’Huillier, M.~Lewenstein, P.~Sali{\`e}res, B.~Carr{\'e}, Theory of high-order harmonic generation by an elliptically polarized laser field, Physical Review A 53~(3) (1996) 1725.

\bibitem{henke1993x}
B.~L. Henke, E.~M. Gullikson, J.~C. Davis, X-ray interactions: photoabsorption, scattering, transmission, and reflection at e= 50-30,000 ev, z= 1-92, Atomic data and nuclear data tables 54~(2) (1993) 181--342.

\bibitem{soldan2001static}
P.~Sold{\'a}n, E.~P. Lee, T.~G. Wright, Static dipole polarizabilities ($\alpha$) and static second hyperpolarizabilities ($\gamma$) of the rare gas atoms (he--rn), Physical Chemistry Chemical Physics 3~(21) (2001) 4661--4666.

\bibitem{gaiser2018polarizability}
C.~Gaiser, B.~Fellmuth, Polarizability of helium, neon, and argon: new perspectives for gas metrology, Physical Review Letters 120~(12) (2018) 123203.

\bibitem{polyanskiy2024refractiveindex}
M.~N. Polyanskiy, Refractiveindex. info database of optical constants, Scientific Data 11~(1) (2024) 94.

\bibitem{kerr1875liv}
J.~Kerr, Liv. a new relation between electricity and light: Dielectrified media birefringent (second paper), The London, Edinburgh, and Dublin Philosophical Magazine and Journal of Science 50~(333) (1875) 446--458.

\bibitem{boyd2008nonlinear}
R.~W. Boyd, A.~L. Gaeta, E.~Giese, Nonlinear optics, in: Springer handbook of atomic, molecular, and optical physics, Springer, 2008, pp. 1097--1110.

\bibitem{perelomov1966ionization}
A.~Perelomov, V.~Popov, M.~Terent’Ev, Ionization of atoms in an alternating electric field, Sov. Phys. JETP 23~(5) (1966) 924--934.

\bibitem{ammosov1986tunnel}
M.~V. Ammosov, N.~B. Delone, V.~P. Krainov, Tunnel ionization of complex atoms and atomic ions in electromagnetic field, in: High intensity laser processes, Vol. 664, SPIE, 1986, pp. 138--141.

\bibitem{derouillat2018smilei}
J.~Derouillat, A.~Beck, F.~P{\'e}rez, T.~Vinci, M.~Chiaramello, A.~Grassi, M.~Fl{\'e}, G.~Bouchard, I.~Plotnikov, N.~Aunai, et~al., Smilei: A collaborative, open-source, multi-purpose particle-in-cell code for plasma simulation, Computer Physics Communications 222 (2018) 351--373.

\bibitem{crank1947practical}
J.~Crank, P.~Nicolson, A practical method for numerical evaluation of solutions of partial differential equations of the heat-conduction type, in: Mathematical proceedings of the Cambridge philosophical society, Vol.~43, Cambridge University Press, 1947, pp. 50--67.

\bibitem{sommerfeld1949partial}
A.~Sommerfeld, Partial differential equations in physics, Vol.~1, Academic press, 1949.

\bibitem{HenkeTable}
BerkleyLab, Henke's form factor tables, \url{https://henke.lbl.gov/optical_constants/asf.html}.

\bibitem{NistTable}
NIST, Nist form factor tables, \url{https://physics.nist.gov/PhysRefData/FFast/html/form.html}.

\bibitem{kulander1990generation}
K.~C. Kulander, B.~W. Shore, Generation of optical harmonics by intense pulses of laser radiation. ii. single-atom spectrum for xenon, Journal of the Optical Society of America B 7~(4) (1990) 502--508.

\bibitem{HarPZenodo}
M.~Pouyez, D.~Mašlárová, P.~Jansson, \href{https://doi.org/10.5281/zenodo.23562049}{Input data for "{HarP}: a simulation tool for high-order harmonic generation and propagation in gases"} (Oct. 2026).
\newblock \href {https://doi.org/10.5281/zenodo.23562049} {\path{doi:10.5281/zenodo.23562049}}.
\newline\urlprefix\url{https://doi.org/10.5281/zenodo.23562049}

\bibitem{l1990propagation}
A.~L’Huillier, X.~Li, L.~Lompr{\'e}, Propagation effects in high-order harmonic generation in rare gases, Journal of the Optical Society of America B 7~(4) (1990) 527--536.

\bibitem{rundquist1998phase}
A.~Rundquist, C.~G. Durfee~III, Z.~Chang, C.~Herne, S.~Backus, M.~M. Murnane, H.~C. Kapteyn, Phase-matched generation of coherent soft x-rays, Science 280~(5368) (1998) 1412--1415.

\bibitem{schnurer1999absorption}
M.~Schn{\"u}rer, Z.~Cheng, M.~Hentschel, G.~Tempea, P.~K{\'a}lm{\'a}n, T.~Brabec, F.~Krausz, Absorption-limited generation of coherent ultrashort soft-x-ray pulses, Physical Review Letters 83~(4) (1999) 722.

\bibitem{constant1999optimizing}
E.~Constant, D.~Garzella, P.~Breger, E.~M{\'e}vel, C.~Dorrer, C.~Le~Blanc, F.~Salin, P.~Agostini, Optimizing high harmonic generation in absorbing gases: Model and experiment, Physical Review Letters 82~(8) (1999) 1668.

\bibitem{kazamias2003global}
S.~Kazamias, D.~Douillet, F.~Weihe, C.~Valentin, A.~Rousse, S.~Sebban, G.~Grillon, F.~Aug{\'e}, D.~Hulin, P.~Balcou, Global optimization of high harmonic generation, Physical review letters 90~(19) (2003) 193901.

\bibitem{he2009spatial}
X.~He, M.~Miranda, J.~Schwenke, O.~Guilbaud, T.~Ruchon, C.~Heyl, E.~Georgiadou, R.~Rakowski, A.~Persson, M.~Gaarde, et~al., Spatial and spectral properties of the high-order harmonic emission in argon for seeding applications, Physical Review A (Atomic, Molecular and Optical Physics) 79~(6) (2009).

\bibitem{rothhardt2014absorption}
J.~Rothhardt, M.~Krebs, S.~H{\"a}drich, S.~Demmler, J.~Limpert, A.~T{\"u}nnermann, Absorption-limited and phase-matched high harmonic generation in the tight focusing regime, New Journal of Physics 16~(3) (2014) 033022.

\bibitem{attosecond_roadmap}
R.~B. Varillas, P.~Agostini, F.~Ardana-Lamas, C.~L. Arnold, D.~Ayuso, M.~Reduzzi, J.~Benda, J.~Biegert, C.~Bourassin-Bouchet, T.~Brabec, C.~Brahms, A.~C. Brown, D.~Busto, J.~Caillat, F.~Calegari, C.~Callegari, S.~Carlstr\"{o}m, Z.~Chang, M.-C. Chen, A.~G. Ciriolo, P.~Corkum, G.~Crippa, R.~d.~Q. Garcia, L.~DiMauro, N.~Dudovich, P.~Eng-Johnsson, D.~Faccialà, P.~Flores, T.~Gadeyne, G.~A. Geloni, C.~Geirger, S.~Gholam-Mirzaei, J.~D. Gorfinkiel, E.~Goulielmakis, M.~Hassan, C.~Hernández-García, P.~Ho, D.~Hui, L.~R. Hutcheson, M.~Ivanov, S.~Kahaly, H.~Kapteyn, N.~Karpowicz, F.~X. K\"{a}rtner, M.~Kling, O.~Kneller, D.~H. Ko, P.~M. Kraus, M.~Kubullek, S.~R. Leone, F.~Lépine, A.~L'Huillier, C.-T. Liao, T.~Linker, A.~G. Lohr, M.~Lucchini, L.~B. Madsen, R.~E. Mainz, B.~Major, J.~P. Marangos, D.~Marco, H.~Marroux, S.~Marshallsay, R.~M. Vázquez, R.~Martín-Hernández, Z.~Mašín, M.~Meyer, F.~M. Moreno, M.~Murnane, D.~M. Neumark, M.~Nisoli, M.~Ossiander, S.~Palakka, S.~Patchkovskii, Z.~Pi, L.~Plaja, J.~Poborska, M.~A.
  Porras, K.~C. Prince, D.~N. Purschke, N.~G. Puskar, G.~M. Rossi, J.~R. Rouxel, T.~Ruchon, P.~Rupprecht, P.~Salières, G.~Sansone, F.~Scheiba, M.~Schultze, B.~Sch\"{u}tte, S.~Serkez, M.~A. Silva-Toledo, O.~Smirnova, S.~Stagira, A.~Trabattoni, J.~C. Travers, I.~Tyulnev, M.~Vacher, G.~Vampa, H.~W. van~der Hart, K.~Varjú, A.-L. Viotti, V.~Vishnoi, M.~Vrakking, V.~Wanie, S.~Witte, F.~Xu, V.~S. Yakovlev, L.~Young, D.~Zhu, C.~Vozzi, \href{https://arxiv.org/abs/2604.15085}{Roadmap on attosecond science} (2026).
\newblock \href {https://doi.org/10.48550/ARXIV.2604.15085} {\path{doi:10.48550/ARXIV.2604.15085}}.
\newline\urlprefix\url{https://arxiv.org/abs/2604.15085}

\bibitem{durfee1999phase}
C.~G. Durfee~III, A.~R. Rundquist, S.~Backus, C.~Herne, M.~M. Murnane, H.~C. Kapteyn, Phase matching of high-order harmonics in hollow waveguides, Physical Review Letters 83~(11) (1999) 2187.

\bibitem{Dominikaspaper}
D.~Maslarova, M.~Pouyez, L.~Hagedorn, P.~Jansson, R.~Weissenbilder, J.~V{\'a}bek, A.~L'Huillier, T.~F\"ul\"op, Physics-informed {Bayesian} optimization of high-order harmonic generation in gases, In preparation for IOP Machine Learning Science and Technology (2026).

\end{thebibliography}
\end{document}